\documentclass[%
 reprint,
 amsmath,amssymb,onecolumn,
 aps,
]{revtex4-2}

\usepackage{graphicx}
\usepackage{dcolumn,float,comment}
\usepackage{bm}

\def \<{\langle}
\def \>{\rangle}
\def \+{\dagger}

\def \CM{\textbf{CM}}

\def \HT{\textbf{HT}}
\def \LRF{\textbf{LRF}}

\renewcommand\<{\left<}
\renewcommand\>{\right>}

\def \bes {\begin{subequations} }
\def \ees {\end{subequations}}
\def \be{\begin{eqnarray}}
\def \ee{\end{eqnarray}}

\usepackage{setspace}
\graphicspath{{Plots/},{}}
\begin{document}

\preprint{APS/123-QED}

\title{The Hypertriton Puzzle in Relativistic Heavy-Ion Collisions}

\author{Thomas Cohen and Maneesha Pradeep\\University of Maryland, College Park}

\date{\today}

\begin{abstract}
The yields of hadrons and light nuclei in relativistic collisions of heavy-nuclei at a center of mass energy of $\sqrt{s_{NN}}\,=\,2.6\, \text{TeV}$ can be described remarkably well by a thermal distribution of an ideal gas of hadrons and light nuclei interacting only via the decay of resonances. Given the particularly small binding energy of hypertritons relative to the temperature describing the yields (about $156\, \text{MeV}$), one might naturally expect hypertrions to dissociate in medium, making the agreement of hypertriton yields with thermal predictions highly puzzling.  The puzzle is compounded by the fact that small binding energy is associated with the large size of the hypertriton.  This size is on a similar scale to the overall size of the fireball and much larger than the length scale over which temperatures in the fireball vary over phenomenologically relevant amounts.   This paper quantifies the tension this effect causes and shows that it is sufficiently large to render the thermal model inconsistent: its natural assumptions are in conflict with its outputs. The possibility that hypertritons are formed at freeze out as compact objects, quark droplets,  that subsequently evolve into hypertritons is considered as a way to resolve the puzzle.  It is noted that beyond making the assumption that compact quark droplets form, additional detailed dynamical assumptions which have not been justified are needed to make the thermal model work.     The issue of why, despite these issues, the hypertriton is well described by a simple statistical description at freeze out is unresolved.  Resolving the hypertriton puzzle is important as it may clarify whether the phenomenological success of the simple thermal model for yields accurately reflects the simple picture of the underlying physics on which it is based. 
\end{abstract}

\maketitle



\section{Introduction}

\subsection{The puzzle}

The generally accepted description of high-energy collisions of heavy nuclei at  RHIC and the LHC has several stages: first an initial preequilibrium period.  Next, to a good approximation, the system is well described as a quark-gluon plasma (QGP) that  evolves in a manner that is well-described via hydrodynamics.  As the QGP evolves, it cools and the systems forms hadrons (and a relatively small number of light nuclei) that eventually freeze out and stream to the detectors\cite{Huovinen:2001cy,Hirano:2002ds,Kolb:2003dz,Muronga:2003ta,Heinz:2005bw,Baier:2006um,Romatschke:2007mq,Song:2007ux,Dusling:2007gi,Baier:2007ix,Luzum:2008cw,Heinz:2009xj,Denicol:2010xn,Schenke:2010rr,Schenke:2011bn,Schenke:2011tv,Bozek:2012en,Heinz:2013th,Gale:2013da,Ryu:2015vwa,Jaiswal:2016hex,Florkowski:2017olj,Busza:2018rrf}. The distribution of observed particles and their properties  are analyzed in order to deduce the properties of the quark-gluon plasma and the thermal history of the fireball.  

Energy densities in these collisions typically reach values larger than the energy density inside of hadrons; this results in the production of  large number of particles, including light nuclei. For example, central collisions of heavy-ions at $\sqrt{s}_{NN}=2.76\, \text{TeV}$ (which is relevant for the discussion in this paper) result in energy densities as large as $12\, \text{GeV\, fm}^{-3}$ in the transverse plane within a pseudo-rapidity between $-0.5$ and $0.5$; this leads to production of tens of thousands of particles \cite{Busza:2018rrf}. These include light nuclei  such as deuteron, helium and hyper-nuclei such as hypertritons as well\cite{ALICE:2015wav}. 

It was reported in Ref.~\cite{Andronic:2017pug}, that the yields of particles produced in the central rapidity region in a heavy-ion collision at $\sqrt{s}=2.76\, \text{MeV}$ at the Large Hadron Collider (LHC) are described quite well by a  statistical hadronization model (SHM) at a chemical freeze out temperature, $T_{\text{F}}\,=\, 156.5\pm 1.5\, \text{MeV}$ and baryon chemical potential, $\mu_B=0.7\pm 3.8\, \text{MeV}$ and a volume $V=5280\pm410 \,\text{fm}^{3}$. The success of this model lies in its agreement with the experimentally observed abundance values that span nine orders of magnitude, and includes strange and non-strange mesons, baryons, hyperons as well as light nuclei and hypernuclei
and their anti-particles.  The model is quite simple: it assumes that the relative yields of hadrons is directly proportional to their density at freeze out as given by a hadron resonance gas model, in which hadrons and nuclei are described by a non-interacting ideal gas of hadrons and nuclei with masses fixed by their values in free space.  The hadron resonance gas properties depend on the temperature and chemical potential at freeze out.  However, for very high energy collisions the chemical potential is essentially zero and can be neglected.

The agreement between the predictions of the model and experimental observations is particularly striking for light nuclei and hypernuclei \cite{ALICE:2015wav,Andronic:2017pug} as these have very low binding energies(B). These binding energies are well below the freeze-out temperature of the model ($156.5\, \text{MeV}$), typically just a few $\text{MeV}$ or in some cases, a few hundred  $\text{keV}$.  For example, the B of deuteron(d) is $2.2
\, \text{MeV}$ \cite{VanDerLeun:1982bhg}.  Even more striking  is the hypertriton (denoted by $H_{\Lambda}$ in this paper), whose B is estimated to be  $0.148\pm\,0.04 \text{MeV}$ \cite{Bertulani:2022vad,STAR:2019wjm,Davis:2005mb,Juric:1973zq,Bohm:1968qkc,Le:2024rkd}
, which is very low relative to the chemical freeze-out temperature of $156.5\, \text{MeV}$. It is hard to understand how such weakly bound nuclei can long survive in such a medium given the interactions of the  nuclear constituents with hadrons of the thermal bath, unless the the thermal bath was extremely dilute. Essentially any collision in the gas with energies on the scale of the freeze out temperature, would be expected to dissociate the nuclei, elastic collisions seem unlikely. The light nuclei are thus sometimes described as ``snowballs in hell''\cite{Braaten:2024cke}.

Thus modeling these particles as being in a non-interacting ideal gas is problematic unless the system is so dilute that collisions are unlikely on relevant time scales \cite{Dashen:1969ep}.  But the density of hadrons in the gas at freeze out is fixed by temperature and is not particularly dilute.    

In fact, there are a number of issues associated with the weak binding of nuclei which make it hard to understand why the statistical hadronization model should be valid for them.  Yet at the phenomenological level the model works well.  As these issues are most acute for the hypertriton which is extremely weakly bound, we focus on it and refer to the  fact the model appears to work well for the hypertriton, despite these issues as the ``hypertriton puzzle'' in relativistic heavy-ion collisions.  One significant part of the puzzle is the fact that weakly bound states in quantum mechanics typically have large physical sizes and this can create a significant tension with the dynamics of the system.  
Various models based on coalescence\cite{Sato:1981ez,Dover:1991zn,Chen:2003qj}  have been studied as alternatives to describe production of deuterons and hypertriton which also seem to predict similar values of yields as the  SHM at $\sqrt{s}_{NN}=2.76\, \text{TeV}$ \cite{Oliinychenko:2018ugs} and at other energies\cite{Schwarzschild:1963zz,Mattiello:1995xg,Nagle:1996vp,Polleri:1997bp,Scheibl:1998tk,Coci:2023daq}. Nevertheless, it is worthwhile to investigate the  production of the hypertriton within the realm of statistical hadronization model (SHM) because of its remarkable phenomenological success despite its simplicity.  Regardless of whether other approaches can describe the same phenomena, the question of why the SHM works as well as it does, is of interest and may offer important clues in understanding the dynamics of heavy ion collisions.

This work focuses on hypertritons.  
The low binding energy ($148 \, \pm 40 \,\text{keV}$) of the hypertriton implies that its wavefunction is spread widely.  If the hypertriton is modeled as a two-body bound state of a $\Lambda$ with a deuteron, the wavefunction has an rms radius of about $10\, \text{fm}$.  The central puzzle associated with hypertriton yields in the SHM is that low binding energy and large physical size of the hypertriton appears to be in contradiction to the assumptions underlying the Statistical Hadronization Model, yet the predicted yields is qualitatively accurate.

In Ref.~\cite{Cai:2019jtk}, it was noted that the time required for the different regions of the hypertriton to be causally connected is long, given the large size of the state.  Thus, causality strongly suggests that whatever mechanism creates the hypertriton, takes a time that is of the order the size of the state (in units where $c=1$) or longer.  This is problematic because of the nature of the hadron resonance gas on which the SHM is based.  The Hadron Resonance Gas (HRG) model estimates are based on an ideal gas of hadrons ({\it i.e. a nearly non interacting gas}), but lower bounds of interactions  rates can be estimated from the assumption that unstable particles such as the $\Delta$ resonance are in equilibrium---a key assumption of the HRG model.  The fact that $\Delta$s decay at a known rate in free space (and in the hot medium, additional mechanisms exist that can increase this rate) gives important information on interaction rates:  detailed balance for equilibrium implies that the rate of the process $N+\pi \rightarrow \Delta$ must compensate for $\Delta$ decays; this, in  turn, sets an upper bound for the lifetime of a nucleon as a distinct particle. An analogous bound lifetime exists for the $\Lambda$.  Since the hypertriton is large and weakly bound, one expects that the reactions turning nucleons and lambdas to other baryons that occur inside the hypertriton, should be similar to that of unbound nucleons and $\Lambda$s. This gives an estimated upper bound for the lifetime of a hypertriton in equilibrium.  

Note, that this is an extremely conservative upper bound: it only includes mechanisms which are inelastic at the hadronic level {\it i.e.} processes where a hadron is converted into other type of hadron---and only a limited set of these.  But, the bound does not include processes that are elastic at the hadronic level {\it i.e.} processes in which kinetic energy and momentum are transferred but the hadrons involved in the collisions retain their identities.  Given the fact that the temperature at chemical freeze out is vastly larger than the binding energy of the hypertriton (and of the deuteron, that is one its components), one would expect that even extremely delicate collisions within the medium would be likely to dissociate the hypertriton.

Ref.~\cite{Cai:2019jtk} noted that this conservative upper bound for the lifetime is approximately one fm, which is an order of magnitude less than the causality lower bound for the formation time.  Thus, given the assumptions of the HRG model, the hypertriton is destroyed long before it is formed:  any prediction of the density of hypertritons by the HRG model at the freeze out  temperature of the SHM  (or above) does not appear to be valid.  This in turn, suggests that justification for the yields as arising from an equilibrated gas of hadrons that freezes out, ought not apply to hypertritons.  Accordingly, it is puzzling that the Statistical Hadronization Model works for hypertritons.  

In contrast, a recent coalescence based approach based on the notion of contact, models these ``snowballs in hell'' as being formed at late times at much lower temperatures and thus over much later times---after kinetic freeze out \cite{Braaten:2024cke}.  But the puzzle of why the predictions of the SHM works qualitatively remains unresolved.

Note that the argument based on the discrepancy between the creation time for a hypertriton based on causality and the hypertriton's lifetime at the chemical freeze out temperature  only depends on properties of the hadron resonance gas and is independent of the details of the dynamical evolution of the fireball.  The purpose of the present paper is to show that even if the implications of that conflict were somehow evaded, the interplay between the dynamics of the fireball and the large size of the hypertriton creates other  puzzles for understanding the success of hypertriton yields in the SHM.

In a dynamical description of the fireball, hydrodynamics taken together with an equation of state, has matter flowing outward and cooling.  Thus, the temperature of the matter inside the fireball has a spatial-temporal distribution with the edge of the fireball at a chemical freeze out temperature.  The natural way to interpret the statistical hadronization model in such a background is that the hadrons (and nuclei) form and equilibrate inside the fireball close to the freeze out surface and flow outward.
One consequence of this, is that the volume seen in the Statistical Hadronization Model, which fixes the total (as opposed to the relative) yields, should be interpreted as the time integral of some shell near the freeze out surface from which equilibrated hadrons emerge.  The shell's width depends on details of the microphysics.   For the SHM to be justified, the temperature throughout the shell needs to be sufficiently close to the freeze out temperature that within it, the hadrons (and nuclei)--which are assumed to be equilibrated have nearly constant density.  However, this is in tension with the large size of the hypertriton which cannot fit into a narrow shell.

This paper quantifies this tension and shows that it is sufficiently large as to render the SHM invalid, in the sense that the results of the model are in contradiction with the models assumptions.  We demonstrate this by modeling both the space-time evolution and the wave-function of the hypertriton.  While both of these are somewhat crude, they are more than adequate to establish that the natural assumptions justifying the SHM do not apply for the hypertriton.  The puzzle is to understand why despite this, the model works phenomenologically for hypertriton yields.

While this paper focuses on how both the spatial and temporal evolution of the fireball cause tensions with the SHM, we are not the first to note that the size of the hypertriton is problematic given the  size of the fireball.  Indeed,  Ref.~\cite{Andronic:2017pug}, in advocating the SHM as a way to decode the phase structure of QCD, also noted that the agreement in the hypertriton yields with the values of the SHM was remarkable, given the large size of the hypertriton relative to the fireball.  Ref.~\cite{Andronic:2017pug} speculated that the agreement might be understood in terms of a production mechanism involving a primary ``colorless quark droplet", compact state with hypertriton quantum numbers that is formed in the fireball at or near freeze out temperatures, which  subsequently evolves into  to the weakly bound hypertriton state seen asymptotically.  As will be discussed in this paper, elementary quantum mechanical considerations suggest that such a scenario does not naturally explain the yields seen in the SHM without further dynamical assumptions. Thus the hypertriton puzzle remains unresolved.

    \subsection{Hydrodynamical model for the evolution of strongly interacting matter}

Before discussing the puzzles involving hypertriton yields, it is useful to review the standard picture of ultra-relativistic heavy ion collisions.

Fundamentally, the dynamics of QCD matter would most appropriately be described in terms of relativistic quantum mechanical evolution of gluon and quark fields, directly from QCD. However, an understanding of this real-time evolution taking into account its highly non-perturbative nature, is well beyond the beyond the state of the art of first-principle calculations on conventional computers due to a sign problem.  Moreover, this is also far beyond the capacity of quantum computers of the near future, given their small size and inability to control noise.  Given this, some approximate effective description of the process is necessary.

If there exists a regime where the relaxation times for the particles or quasi-particles in the system to equilibrate are far shorter than the relevant timescales, such the inverse temperature,
it is plausible to approximate the full quantum evolution by a hydrodynamic description. If this is the case, then the variation of the thermodynamic properties, such as the conserved densities can be neglected over length scales where quantum mechanical description of the structure is inevitable. This seems reasonable for description of protons and pions whose microscopic structure may be important only for length scales of the order of $1\, \text{fm}$. However this becomes problematic for hypertriton due to its very large size. We  quantify this tension in this paper.

Thermalization in heavy-ion collisions has important consequences. It is hoped that heavy-ion collisions can be used to study the thermodynamics of Quantum Chromodynamics, the fundamental theory which governs the behavior of quark-gluon plasma, as well as the hadrons, that they get confined into at low temperatures. One of the grandeur visions of the heavy-ion community is to map out the phase diagram of QCD by colliding the heavy-ions at varying center of mass energies. This vision relies on the assumption that the particle production in heavy-ion collisions can be described by an appropriate close-to-equilibrium description of the strongly-interacting matter at a certain stage in the expansion of the system post-collision, called freeze-out where the particle distributions get frozen\cite{An:2021wof}.

Experimental observations suggest that the quark-gluon plasma formed shortly after the collision achieves local equilibration within $1\, \text{fm/c}$\cite{Heinz:2001xi,Niemi:2015qia}. The system thermalizes forming a dense and hot medium that is well-described directly in terms of quarks and gluons; the subsequent evolution can be described by hydrodynamics. The hydrodynamic equations are simply conservation equations for the energy-momentum and charges. The only QCD inputs to the model are the equation of state(EoS), i.e a relationship between the pressure and energy and number density and the viscosity and diffusion coefficients and their dependence on temperature and chemical potential. 
As the system expands and becomes more and more dilute, hydrodynamics breaks down and becomes inapplicable.  This occurs when the average inter-particle spacing for the particles or quasiparticles associated with the relevant degrees of freedom become comparable to or larger than  than the typical distance scale associated with scattering.  As the inter-particle distances get still larger, inelastic processes that can change species type  eventually cease to happen. Later, the same happens to the elastic processes that maintain kinetic equilibrium. The separation of these stages are commonly referred to as chemical and kinetic freeze-out respectively. All these happen at times $ \sim 10-15\,\text{fm/c}$ after the collision in the lab-frame. 
 
In the SHM, the total yields of the stable particles observed at the detectors by this remarkably simple picture are therefore determined at the chemical freeze-out and their momentum distribution can be traced back to the kinetic freeze-out.  The Statistical Hadronization Model  assumes that shortly before  chemical freeze-out, the system was well described by a nearly ideal hadron Resonance gas in near-thermal equilibrium. 

    The break-up of this paper as follows. In Section.~(\ref{model}), we describe the hydrodynamical model for the evolution of the fireball untill freeze-out and the quantum mechanical model we use for the hypertriton in our analysis. In Section.~(\ref{main}), we utilise the fireball profile from the hydrodynamical simulation at various instants during its evolution, along with the quantum mechanical model for the hypertriton, to obtain the probability that the hypertriton can be ``fit" inside the fireball. We further discuss the distribution of the effective temperatures felt by the constituents of the hypertriton. Based on our analysis, we conclude that the likelihood that the SHM model is consistent with a quantum-mechanical description of hypertriton is typically much less than $25\,\%$. We conclude in Section.~(\ref{last}) with a few comments on whether compact colorless quark droplets at freeze-out can evolve into the hypertriton state, in light of the analysis presented in this paper.

\section{Hypertriton inside the fireball}
\label{model}
\subsection{Thermal history of the fireball until chemical  freeze-out}

\label{hydrodynamics}

In this section, we will describe a simple hydrodynamical model that will be used to determine the size of the fireball when it freezes-out into a gas of hadrons and nuclei. The evolution of the energy density, and the fluid four-velocities in our simulation is determined using the standard MIS second order
hydrodynamic equations as implemented in the publicly available
VH$1 + 1$ hydrodynamic code~\cite{Baier:2006gy, Baier:2006um,
Romatschke:2007jx} and further modified in Ref.~\cite{Rajagopal:2019xwg}.  The Equation of State (EoS) used in the
simulation was introduced in Ref.~\cite{Rajagopal:2019xwg} and also used in Ref.~\cite{Pradeep:2022mkf}. Appendix.~(\ref{appendeos}) shows plots of some of the thermodynamic quantities derived from this EoS.  More details about the EoS are in the references. This EoS allows for the inclusion of the effect of a critical point, however, we have turned this off for our calculation.

The shear viscosity to entropy density ratio is set to $\eta/s=0.12$ throughout, 
and the hydrodynamic equations are numerically solved using a spatial (radial) lattice with 1024 points spaced by $0.0123$~fm and a time step of $0.005$~fm. The bulk viscosity is set to zero. The fireball is assumed to equilibrate sufficiently that hydrodynamics is applicable early in the collision,  typically taken to be at time of approximately $\tau\equiv\tau_{I}=1\, \text{fm}$ in the lab frame\cite{Berges:2020fwq}. For our analysis, we use hydrodynamic simulations that are initialized at $\tau_I=1\, \text{fm}$ with an initial
central temperature of 330 MeV, 400 MeV and 500 MeV to describe the dependence of results on the value of the initial central temperature.
The closest guidance for this value is obtained from Ref.~\cite{Bozek:2010wt}, which obtains the temperature at $\tau_{I}=1\,\text{fm}$ for a $2+1$ hydro simulation that simulates a Pb-Pb collision with center of mass energy, $\sqrt{s_{NN}}\,=\,2.76\,\text{TeV}$ to be $480\, \text{MeV}$.  The
radial flow and the viscous part of the stress-energy tensor are set to zero initially  at
$\tau=\tau_I$. 
A standard Glauber model radial profile 
for the energy density at
$\tau=\tau_i$, with the Wood-Saxon parameters, characteristic radius, $R\,=\,6.4\, \text{fm}$ and $a\,=\, 0.54\, \text{fm}$ has been assumed, following Ref.~\cite{Baier:2006gy}.

Of course, more sophisticated hydrodynamic models with anisotropies exist for describing various properties such as elliptic flows and centrality dependence of particle multiplicities at $\sqrt{s_{NN}}\,=\, 2.76\, \text{TeV}$ \cite{Bozek:2012qs,Alqahtani:2017tnq}.  We are using this very simple model as our goals are the study of qualitative features that simple models such as this should capture.  Moreover, a simple model has fewer parameters than more sophisticated ones and the conclusions that we reach should not depend on the details of those extra parameters. 

In our model, the flow is azimuthally symmetric and longitudinally boost-invariant. This is sufficient for our purposes, since we are concerned with hyper-nuclei yield in central collisions. 
We assume vanishing net-baryon density which is reasonable for Pb-Pb central collisions at high center of mass energies such as $2.76\, \text{TeV}$ per nucleon, where the baryon chemical potential is very small and can be neglected relative to the temperature at freeze-out.  The likelihood of the hypertriton fitting inside the fireball, is in general, expected to decrease as the hydrodynamic description deviates from an ideal description with no viscosities. We studied numerically the effect of different values of $\eta/s$ and find the effects to be small. 

The chemical freeze-out hypersurface is considered to be an isothermal surface at $T=156.5\, \text{MeV}$. The azimuthal symmetry and the longitudinal boost invariance defines a cylindrical boundary for a hadron at mid-rapidity. Note, that the freeze-out is not isochronous in the lab frame due to the non-trivial radial expansion, meaning fluid cells at different radii freeze-out at different times in the lab frame.   We have chosen $T=156.5\, \text{MeV}$ as the freeze out temperature to match the chemical freeze out temperature of the SHM at the center of mass energy, $\sqrt{s_{NN}}=2.76\,\text{TeV}$. We estimate the radial sizes of the fireball, determined by the location of the freeze-out hypersurface, as a function of the time, as the edges of the fireball gradually but continuously freeze out.

\begin{figure}[H]
  \centering
\begin{minipage}{0.45\textwidth}
  \includegraphics[width=\textwidth]{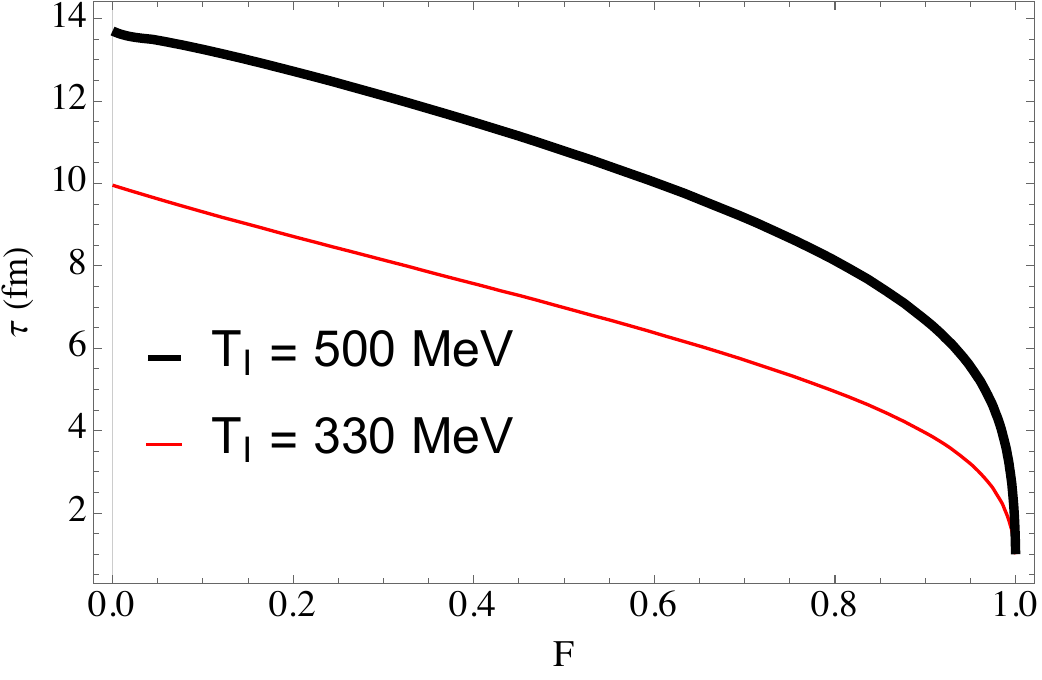}
  \end{minipage}
  \begin{minipage}{0.45\textwidth}
  \includegraphics[width=\textwidth]{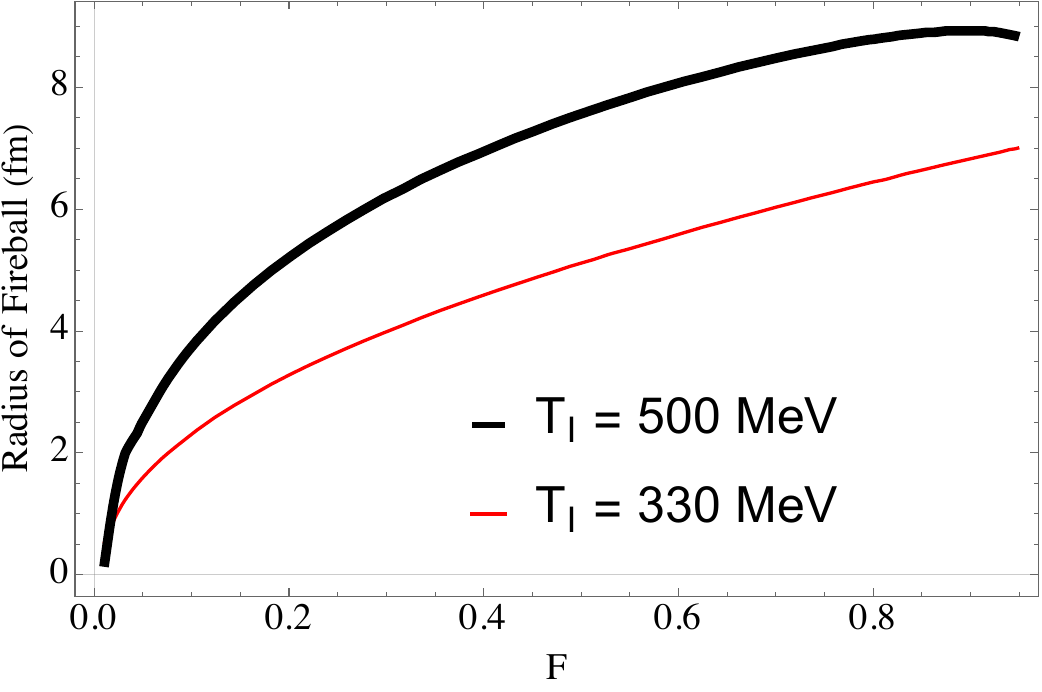}
  \end{minipage}
  \caption{The time and the radius in the center of mass frame of collision when $F$ fraction of the fireball is remaining}
\label{Fig:RandtauVsF}
\end{figure}

Matter is constantly flowing outward in this hydrodynamic description.  Eventually the matter flows beyond the freeze-out hypersurface.   In our model, we consider matter to be within the fireball until it passes the freeze-out hypersurface.  Thus, although matter is flowing outward, the size of the fireball--parameterized in terms of the radius in the lab frame of  freeze-out hypersurface is decreasing for most, or all, of the evolution of the fireball. For the purposes of this paper, rather than parameterizing the fireball directly in terms of this radius,  it is useful to parameterize the evolution of the fireball in terms of the fraction of the energy that remains inside the fireball relative to the energy in the fireball at the start
of the  hydrodynamic evolution (taken to be  $\tau_I=1\, \text{fm})$.  Accordingly, in the remainder  of the paper, we will use the fraction of the fireball remaining, denoted by $F$ as a proxy for the time that has elapsed. In Fig.~(\ref{Fig:RandtauVsF}), we show the radius of the fireball and the time since the start of the collision as a function of the fraction of energy remaining in the fireball.

\begin{figure}[H]
  \centering
 \includegraphics[width=0.6\textwidth]{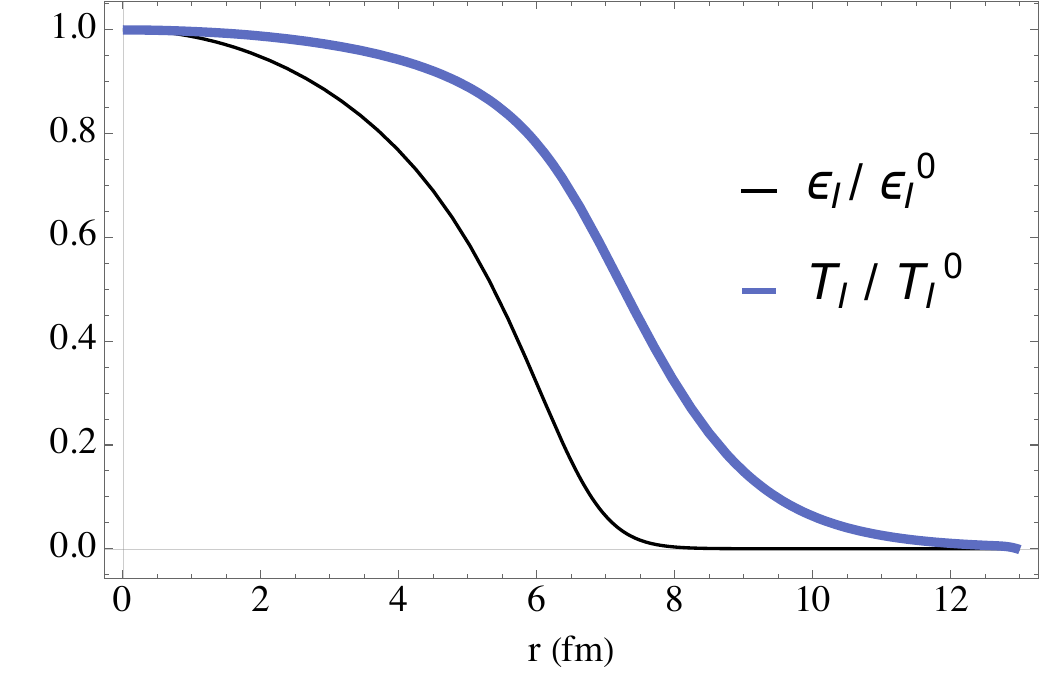}
    \caption{Temperature and energy density profile of the fireball relative to their magnitudes at the center of the fireball at $\tau=1\, \text{fm}$ with a central temperature of 330 MeV.}
\label{IntialProfile}
\end{figure}

\subsection{Hypertriton wavefunction in vacuum}

\label{wavefunction}

Another key to the analysis is the description of the hypertriton.  One can view a hypertriton as a bound state of $\Lambda$ and a deuteron.  In this section, we describe the wavefunction for the hypertriton state in terms of a wavefunction for a $\Lambda-d$ bound system.    While one might think that there is a need to describe this in medium, we consider the wave function in vacuum since the hadron resonance gas model assumes a gas that is sufficiently dilute that properties of its constituents are given by their vacuum properties.

An analysis of the existing data suggests that the binding energy between $\Lambda$ and $d$ in a hypertriton is exceptionally small.  Infact, it is so small that it is hard to measure accurately; the current estimate is a binding energy of only $148\pm 40\, \text{keV}$ \cite{Bertulani:2022vad}.  This is very low relative to the binding energy of deuteron, which is about $2.2 \, \text{MeV}$ \cite{VanDerLeun:1982bhg}, which is itself very small on the scale of typical nuclear binding. The mean separation between the $\Lambda$ and $d$ in the hypertriton system can be deduced from this value of binding energy to be about $\sim 10 \, \text{fm}$ \cite{Bertulani:2022vad}. 
That qualitative size follows from the following elementary consideration: using general quantum mechanical considerations probability density for finding $\Lambda$ and the deuteron within the hypertriton at a distance, $r$, at long distances (longer than the range of the interaction between the $\Lambda$ and deuteron) is given  by
\begin{equation}
\rho(r) = \rho_0 \exp \left (-\sqrt{8 \mu B} r \right) \label{Eq:long}
\end{equation}
where $\mu$ is the reduced mass of the $\Lambda$-deuteron system.  Since $B$ is extremely low, it means that the probability distribution has a very long-ranged tail; 
$\rho_0$ is fixed by the integrated probability that the distance between the deuteron and the $\Lambda$ is beyond the range of their interaction.   Moreover, it is easy to show that as $B \rightarrow 0$, $\rho_0 \rightarrow \sqrt{8 \mu B}(1+l_0 \sqrt{8 \mu B})$, (where $l_0$ is a characteristic length scale associated with the potential)  which implies that as $B \rightarrow 0$, the fraction of the probability in the regime where Eq.~(\ref{Eq:long}) holds approaches unity.  Plugging the small binding energy and neglecting the short-distance part of the wave function as unimportant (as it is for sufficiently small $B$) implies  that ``typical'' quantities characterizing the distances for the size are $\sim 10 \, \text{fm}$.  

For our purposes a highly accurate description of the hypertriton state is not required.  However, it is necessary to have a somewhat better description than simply stating that it has a size of about 10 fm.  Two features are important.   The first is that while the description of the system as the bound state of a $\Lambda$ and deuteron is basically adequate for our purposes, the deuteron itself is fairly large (reflecting the fact that it too is weakly bound given typical nuclear scales) and its size is not negligible.  The second is that while the short-distance part of the $\Lambda$ deuteron wave function has negligible probability in the limit of arbitrarily weak binding, even with the very small binding of the hypertriton it is not completely negligible due the large reduced mass and the unfortunate factor of $\sqrt{8}$. 

Thus we will model the hypertriton as a two-body bound state between the center of mass of the a deuteron and a $\Lambda$.  This will be modeled using a simple potential model.  Folded into this is a distribution of proton and neutrons relative to the deuteron bound state (given by a deuteron wave function).

The potential model for the hypertriton is not well constrained. 
High quality scattering data for very low energy $\Lambda$-deuteron scattering is not available, precluding the establishment of a high quality $\Lambda$-d potential capable of producing a reliable wave function of the bound state.  Moreover, the break-up threshold for deuteron is quite low, so such a potential model would have limited validity in any case.   Fortunately, for the present purposes, a crude description of the hypertriton wave function should be adequate.  What principally matters for our purposes is a model that qualitatively reproduces the overall size of the hypertriton and a reasonable estimate for how much of the wave function is at short distance, relative to the thermal gradient length scales within the fireball.

Accordingly, we will use a simple form, modeling  the interaction between the $\Lambda$ and deuteron  by a finite square well-potential, specified by the depth of the potential and its finite radial extent, $c$. The parameter $c$ is taken to be $2.5\, \text{fm}$, a value that seems phenomenologically plausible  and which has been used previously in the literature \cite{Bertulani:2022vad}; the potential depth has been chosen to reproduce the binding energy (using $B  = 148 \, \text{keV}$ ).  

\begin{figure}[H]
  \centering
\begin{minipage}{0.45\textwidth}
  \includegraphics[width=\textwidth]{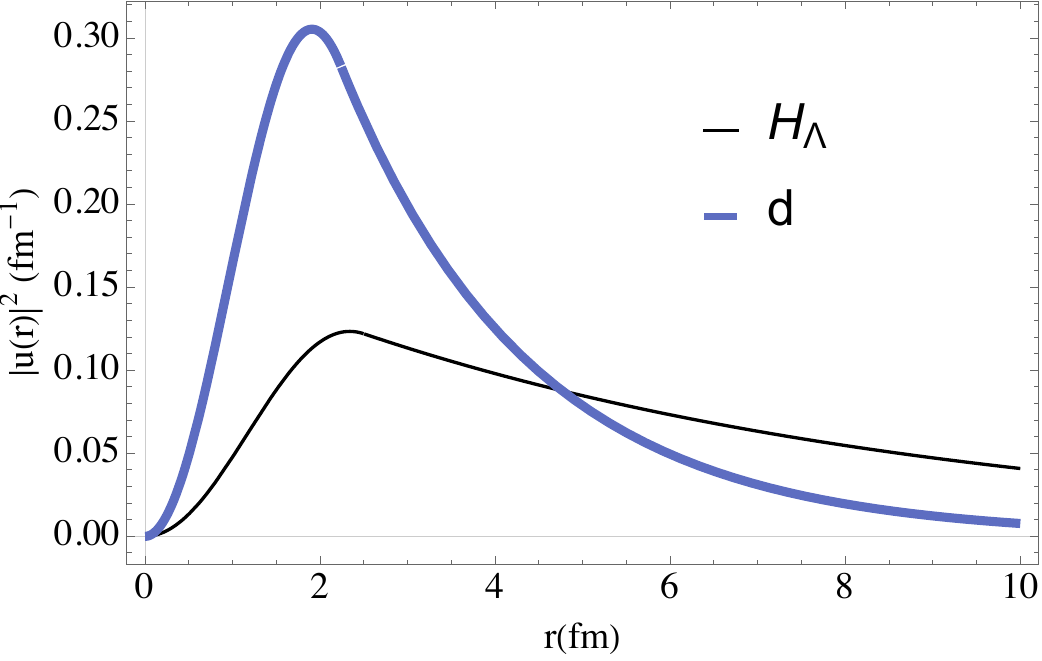}
  \end{minipage}
  \begin{minipage}{0.45\textwidth}
  \includegraphics[width=\textwidth]{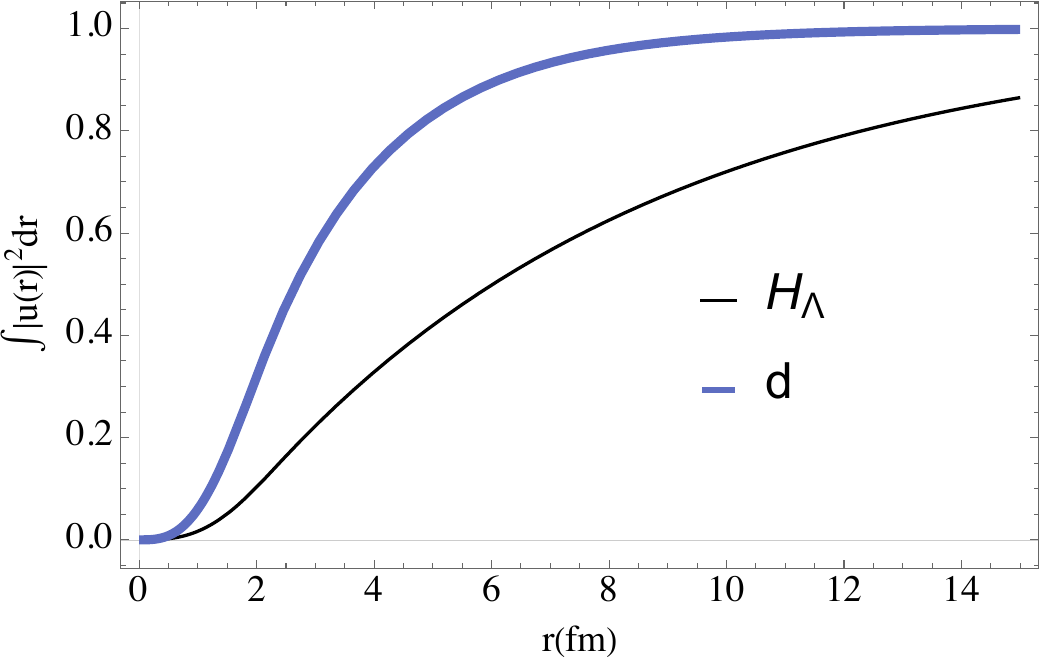}
  \end{minipage}
  \caption{The square of the radial wavefunction divided by the radial separation and the integrated probability density for the deuteron -Lambda system (denoted $H_\Lambda$)  and  the deuteron (denoted $d$) as a function of the distance between their constituents.}
\label{Fig:Wavefunction}
\end{figure}

In principle, we could use a deuteron wave function based on sophisticated potential model that fits the nucleon-nucleon scattering data.  However, there is no reason to do so, given the uncertainties and crudeness in the deuteron-$\Lambda$ wave function. Our results are far less sensitive to the details of the distribution of nucleons in the deuteron than to the deuteron-$\Lambda$ wave function (principally because the deuteron is far smaller). Hence, it suffices to neglect the $D-$wave components in the deuteron and model the s-wave proton-neutron potential also as a square well (with parameters fit to reproduce the deuteron charge radius and binding energy). The finite radial extent of the deuteron finite square well is taken to be $2.24\, \text{fm}$ so as to reproduce the deuteron charge radius of about $2.13\, \text{fm}$\cite{Hernandez:2017mof} and the depth of the potential is chosen to get a binding energy of $2.23\, \text{MeV}$\cite{VanDerLeun:1982bhg}. The  wave functions are shown in Fig.~(\ref{Fig:Wavefunction}). 

One may worry  that the potential of the hypertriton and deuteron would be substantially modified inside the strongly interacting medium composed of a QGP or hadrons---or indeed that a simple potential model description becomes inappropriate.  There is scattering in the medium which can modify the bound states. However, this study is to test the assumptions of the Statistical Hadronization Model and  that model assumes that ordinary hadrons (i.e. hadrons that have their free space properties) leave the fireball at chemical freeze out. Therefore, the simple quantum-mechanical model we outlined above is adequate.

\section{Probability that the hadronic constituents of the hypertriton are inside the fireball}
\label{main}

This section is the crux of the paper. Here, we estimate the likelihood that the hadronic constituents of the hypertriton lie within the instantaneous freeze-out hyper-surface of the fireball when the center of mass is inside the fireball. 

\begin{figure}[H]
  \centering
\begin{minipage}{0.45\textwidth}
  \includegraphics[width=\textwidth]{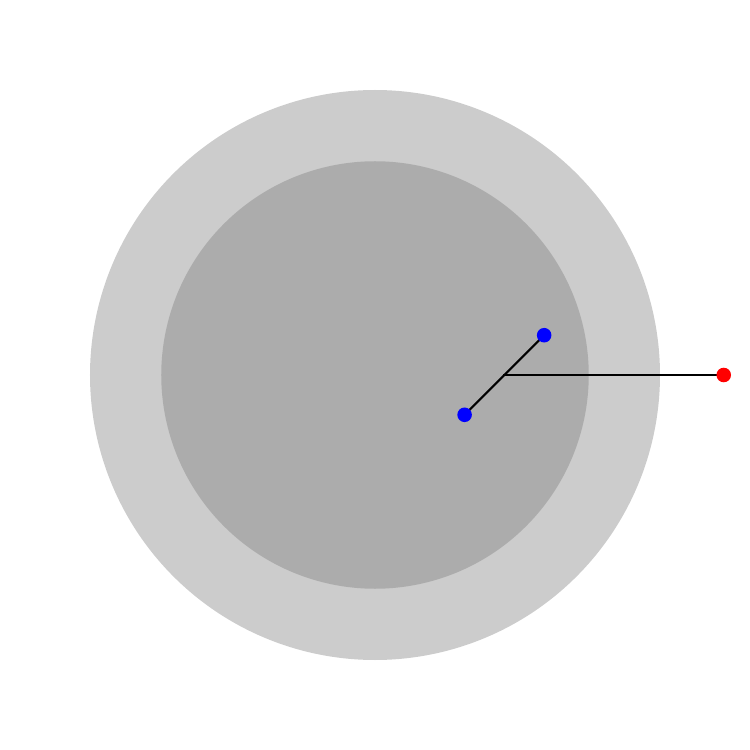}
  \end{minipage}
  \begin{minipage}{0.45\textwidth}
  \includegraphics[width=\textwidth]{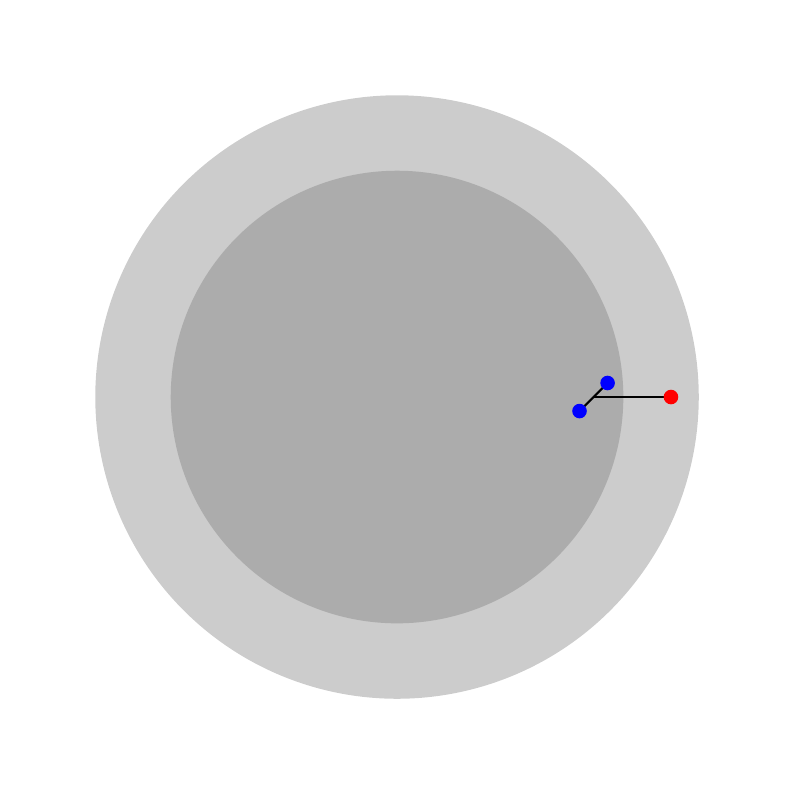}
  \end{minipage}
  \caption{Illustration of  scenarios in which the representative hypertriton whose size is determined by the semi-classical model (described in the text) lies either outside (left) or inside (right) of the fireball.The hypertriton in our semi-classical model is represented by the combination of rods whose lengths are determined by the quantum mechanical wavefunction of deuteron and hypertriton which are explained in Section.~(\ref{wavefunction}) and shown in Fig.~(\ref{Fig:Wavefunction}). The blue points represent the nucleons inside the deuteron and the red point represents $\Lambda$.
  The center of mass of the deuteron-$\Lambda$ system is taken to be at a certain temperature denoted by $T_{\HT}$. The gray region defines the size of the fireball. The region with temperatures greater than $T_{\HT}$ are shown with a darker gray shade.}
\label{Fig:HT}
\end{figure}

Imagine a classical picture in which the deuteron is described by a rigid rod connecting the proton and neutron and the  hypertriton contains another rigid rod  connecting the center of mass of the proton-neutron rod to a third point object, $\Lambda$ (see Fig.~(\ref{Fig:HT})). 
Quantum mechanics enters the analysis by providing the probability distribution for the separation vectors between the objects, which refers to the lengths of the rods and the angles between them. Quantum mechanics renders the characteristic size of the hypertriton a probabilistic quantity. The center of mass of  the hypertriton is also in general moving at a non-zero velocity with respect to the center of the fireball. We ask what fraction of the time, the characteristic size of the hypertriton is small enough to fit all three of its hadronic constituents inside the fireball, when the center-of-mass of the hypertriton is at some given position inside the fireball and moving with some velocity (which causes Lorentz contraction).

The integrated radial wave function of the hypertriton gives the probability of enclosing the separation vector between the $\Lambda$
and $d$ within a specified radius. For the potential described in the previous section, the mean radial separation and the root-mean-squared radial separation between the $\Lambda$ and $d$ is about 8.1 fm and 10.6 fm. The deuteron itself is a composite object of a proton and a neutron separated radially from each other by a mean distance of $3.275\, \text{fm}$ and a root-mean-squared distance of $3.95\, \text{fm}$. We estimate the quantum mechanical likelihood of finding all the hadronic constituents of the hypertriton state inside the hyper-surface that defines the boundary of the hydrodynamic phase at any instant  (instantaneous freeze-out hypersurface) in the center of collision frame or lab frame, denoted as $\CM$ frame.  For this analysis, we consider the  proton, neutron and $\Lambda$ as point particles; doing so gives a conservative estimate. 

Since lengths are frame dependent due to Lorentz contraction effects, and there are various frames of reference involved in the problem, it is useful to denote the relevant frames. The first, is of course, the $\CM$ frame of the fireball fixed by the center of mass of the collision itself. There is a second frame, which is at rest with respect to the fluid (which is moving outward in the rest frame of the fireball), which we denote by $\LRF$, abbreviating local rest frame.  The hypertriton moves due to thermal effects relative to the fluid and its  rest frame  will be denoted by $\HT$. We assume that the hypertritons are produced with a momentum distribution specified by an ideal Boltzmann gas of particles in the $\LRF$---as is assumed in the SHM. The associated temperature, we take to be the temperature at the center of the hypertriton, which by the hypothesis of the SHM is close to the freeze-out temperature.  It is not necessarily exactly equal to the freeze-out temperature since one expects hadrons and nuclei to have a point of last species-changing collision near, not precisely at the freeze out surface but slightly inside it. We call this temperature at that point $T_{\HT}$.  Clearly, $T_{\HT}>T_{f}$.

The Lorentz contraction due to the thermal distribution in velocities and the non-zero radial velocity of the fluid reduces the effective size of the hypertriton in the $\CM$ frame. We perform Lorentz boosts that take the spatial ends of the separation vectors between the $\Lambda$ and the deuteron and those between proton and neutron at different times in the local rest frame of the hypertriton (denoted by $\HT$), to the local rest-frame of the fluid (denoted by $\LRF$) and consecutively to the $\CM$ frame. The time separations of the observations of the ends of the separation vectors in the $\HT$ frame are then fixed so as to make them spatial in the $\CM$ frame. This procedure gives us the Lorentz contracted separation vector at equal times in the $\HT$ frame.

For a hypertriton with center of mass located at a transverse radial coordinate of $R_{\HT}$ and mid-rapidity to lie within the fireball, the longitudinal coordinate of the hadronic constituents should lie within $(-\tau_F,\tau_F)$ where $\tau_F$ is the instantaneous time at freeze out in the lab frame and the transverse radial coordinate should be less than $R_F$, the transverse radial distance of the instantaneous freeze-out hypersurface from the origin. The temperature at the center of mass of the hypertriton specifies the transverse radial distance from origin $R_{\HT}$. The probability of fitting the hypertriton inside the freeze-out hypersurface increases with increasing $R_F-R_{\HT}$ or equivalently with increasing $T_{\HT}-T_{F}$. While there is no clear phase separation between QGP and HRG regime at low values of baryon chemical potential, it is clearly unreasonable  phenomenologically for hypertritons to exist in a medium in which the temperature at the center of mass of the hypertriton was larger than $15 \, \text{MeV}$ above the crossover temperature, which in this case, coincides quite well with the freeze-out temperature of $156.5\, \text{MeV}$\cite{HotQCD:2018pds,Borsanyi:2020fev,Ding:2024sux}.

\begin{figure}[H]
  \centering
  \begin{minipage}{0.45\textwidth}
  \includegraphics[width=\textwidth]{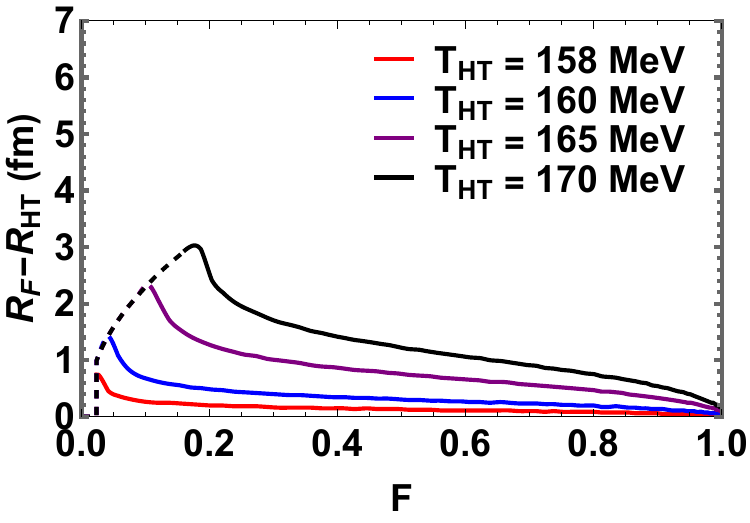}
  \end{minipage}
  \begin{minipage}{0.45\textwidth}
  \includegraphics[width=\textwidth]{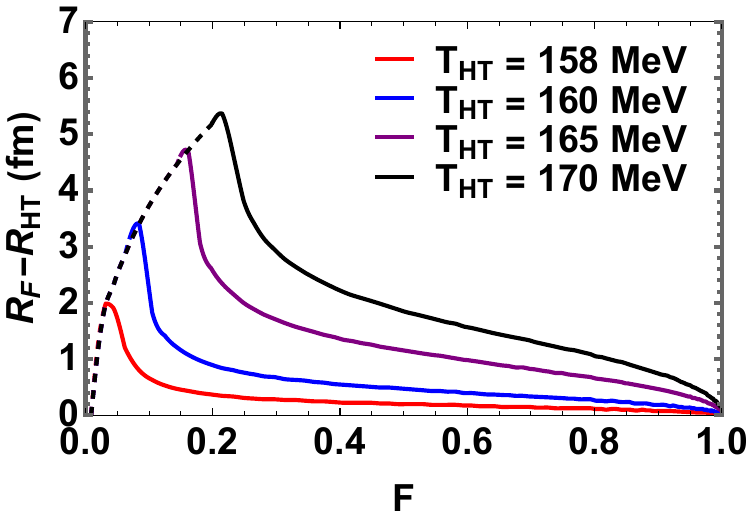}
  \end{minipage}
  \caption{For the left and the right plots, the radial separation between the isothermal surfaces defined by $T_{HT}$ and $T_F=156.5\, \text{MeV}$ as a function of the fraction of remaining fireball (denoted by $F$), from hydrodynamic simulations with initial central temperatures $T_{I}=330\, \text{MeV}$ and $T_{I}=500\, \text{MeV}$ respectively are shown by thick lines. Notice that at sufficiently late times, or equivalently when the fraction of initial energy within the fireball is sufficiently low, there are no regions within the fireball with temperatures greater than or equal to $T_{\HT}$. At these times, the dashed lines show the transverse radius of the remaining fireball. $T_{\HT}$, defined in the text is the temperature at the center of mass of the hypertriton. Therefore, the dashed regions correspond to scenarios in which the hypertriton is at the origin in the collision frame.}
\label{Fig:SeparationVsFMWithtLC}
\end{figure}

For our purposes, the phenomenologically relevant distance is the distance between the chemical freeze out hypersurface and the isotherm at the center of mass of a hypertriton that freezes out without further interaction.  We denote the transverse radial separation between these two isotherms by $R_F-R_{\HT}$. This is shown in Fig.~(\ref{Fig:SeparationVsFMWithtLC}). The significance lies in the fact that the SHM assumes that the temperature of the hypertriton that fixes its yield, must be sufficiently close to the freeze out temperature that the density (and hence the relative  yield) is, to a good approximation, given by the freeze out temperature .  At the same time, for the model to be consistent, $R_F-R_{\HT}$ needs to be large enough to accommodate a significant fraction of the hypertriton's components. We will choose reasonable values for $T_{\HT}$ based on the considerations in the last paragraph and subsequently fix $R_{\HT}$ and determine at various instants of freeze-out, what fraction of times the hadronic constituents of the hypertriton lie within the fireball. 

Of course, $R_F-R_{\HT}$ (for a given $T_{\HT}$) depends on the variable parameters in the hydro model. In our case, these are the initial conditions of the fireball at $\tau=\tau_{I}=1\, \text{fm}$ where the hydrodynamic evolution sets in and the rate of evolution of the energy density, which depends on the viscosities in the medium. Here we consider two initial values of the central temperatures, namely $T_{I}=330\, \text{MeV}$ and $T_{I}=500\, \text{MeV}$ to illustrate the effect of the initial temperature on the analysis. We choose a reasonable estimate of $\eta/s=0.12$ for our analysis.  These are shown in Fig.~(\ref{Fig:SeparationVsFMWithtLC}). The transverse radius of the fireball, defined by the location of the freeze-out hypersurface shrinks as a function of time in the $\CM$ frame while the system expands along the longitudinal direction. The transverse radial separation between the the freeze-out hypersurface and the isothermal surface defined by $T=T_{\HT}$ is shown as a function of the fraction of matter remaining inside the fireball, denoted by $F$, in Fig.~(\ref{Fig:SeparationVsFMWithtLC}), for different values of $T_{\HT}$. Here, $F$ is used as a proxy for the instantaneous time in the $\CM$ frame as it is the quantity of relevance in our analysis.  By construction, $F$  decreases with time. 
\footnote{ $F$ as a function of time is determined as follow: At $\tau=\tau_I$,when the hydrodynamic evolution sets in, the radius within which a certain fraction of the total energy density lies is determined. Following the characteristic trajectories of the fluid elements at these radial distances until they freeze-out, the time at which a certain fraction of fireball is remaining is calculated.}

\begin{figure}[H]
  \centering
  \begin{minipage}{0.45\textwidth}
\includegraphics[width=\textwidth]{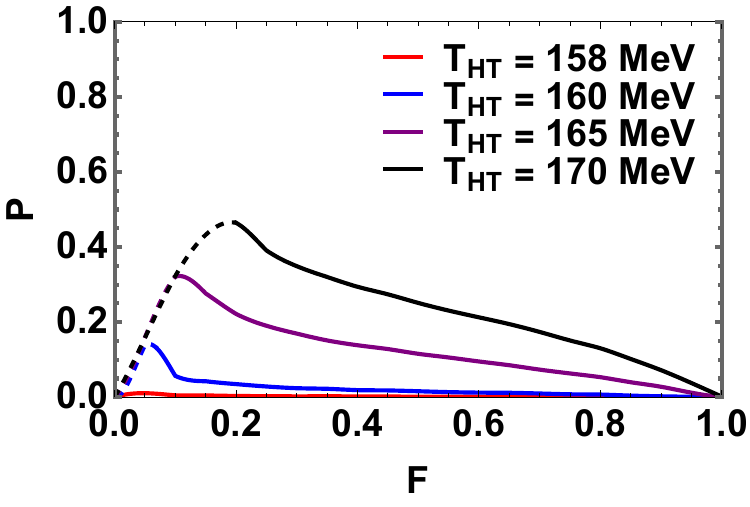}
\end{minipage}
\begin{minipage}{0.45\textwidth}
\includegraphics[width=\textwidth]{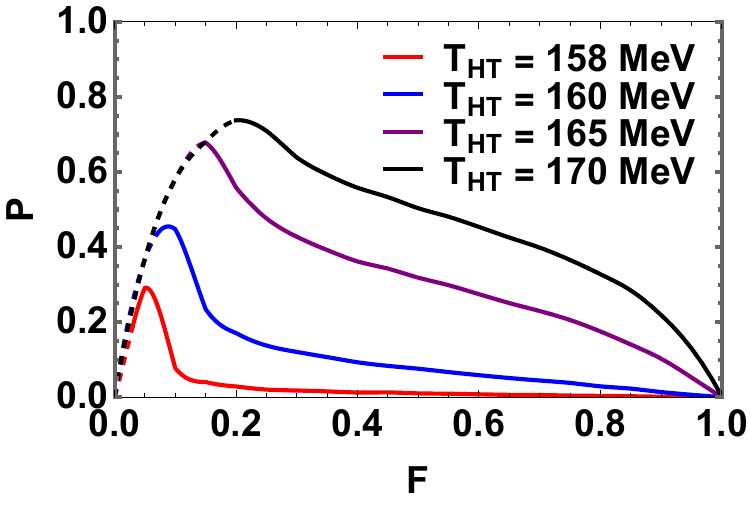}
\end{minipage}
  \caption{The instantaneous probability that the constituents of the hypertriton, the $\Lambda$, the proton and the neutron are inside the freeze-out boundary specified by the condition $T_F=156.5\, \text{MeV}$ with $\eta/s=0.12$ and  initial central temperature, $T_{I}=330\, \text{MeV}$(left) and,  $T_{I}=500\, \text{MeV}$ (right). The fraction of initial energy density remaining inside the fireball, denoted by $F$ is used as a proxy for the time since collision.}
\label{Fig:PVsFWithLC}
\end{figure}

As an intermediate step  we compute and plot in Fig.~(\ref{Fig:PVsFWithLC}), the probability that the hypertriton fits inside the fireball (in the sense that  all three components--the proton, the neutron and $\Lambda$---are all inside the fireball) as a function of $F$, i.e the fraction of the initial energy still inside the fireball. This is done while fixing the center of mass of the hypertriton at different transverse radial distances from the origin on the mid-rapidity plane. 
After a certain time the fireball doesn't have regions hotter than $T_{\HT}$; in these situations we consider the center of mass of the hypertriton to be at the center of the fireball. These are shown by dashed lines in Fig.~(\ref{Fig:PVsFWithLC}).  We calculate the probability numerically via the generation of an ensemble with probabilities  given by our simple quantum mechanical model.
Each  hypertriton is given a  velocity  drawn from a Boltzmann distribution in the $\LRF$ of the fluid. 
We illustrate this for hypertriton temperatures between 158 MeV and 170 MeV.  We consider the 170 MeV surface to be an extremely conservative upper bound for reasonable temperatures for the shell in which hypertritons are equilibrated.  This is well beyond the crossover temperature seen on the lattice\cite{HotQCD:2018pds,Borsanyi:2020fev,Ding:2024sux}. The lattice data suggest that the HRG is breaking down at temperatures that high\cite{HotQCD:2014kol,Bazavov:2017dus}.  It is highly probable that hypertritons would dissolve by the time hadrons become sufficiently strongly interacting that the HRG deviates from the lattice.  Moreover, even if it is meaningful to assert  that hypertritons exist at 170 MeV, the density as calculated in the hadron resonance gas (and hence the predicted yield)  is a factor of five larger than at the nominal freeze out temperature of 156.5 MeV, leading to a phenomenological discrepancy within the SHM.

\begin{figure}[H]
  \centering
\includegraphics[width=0.7\textwidth]{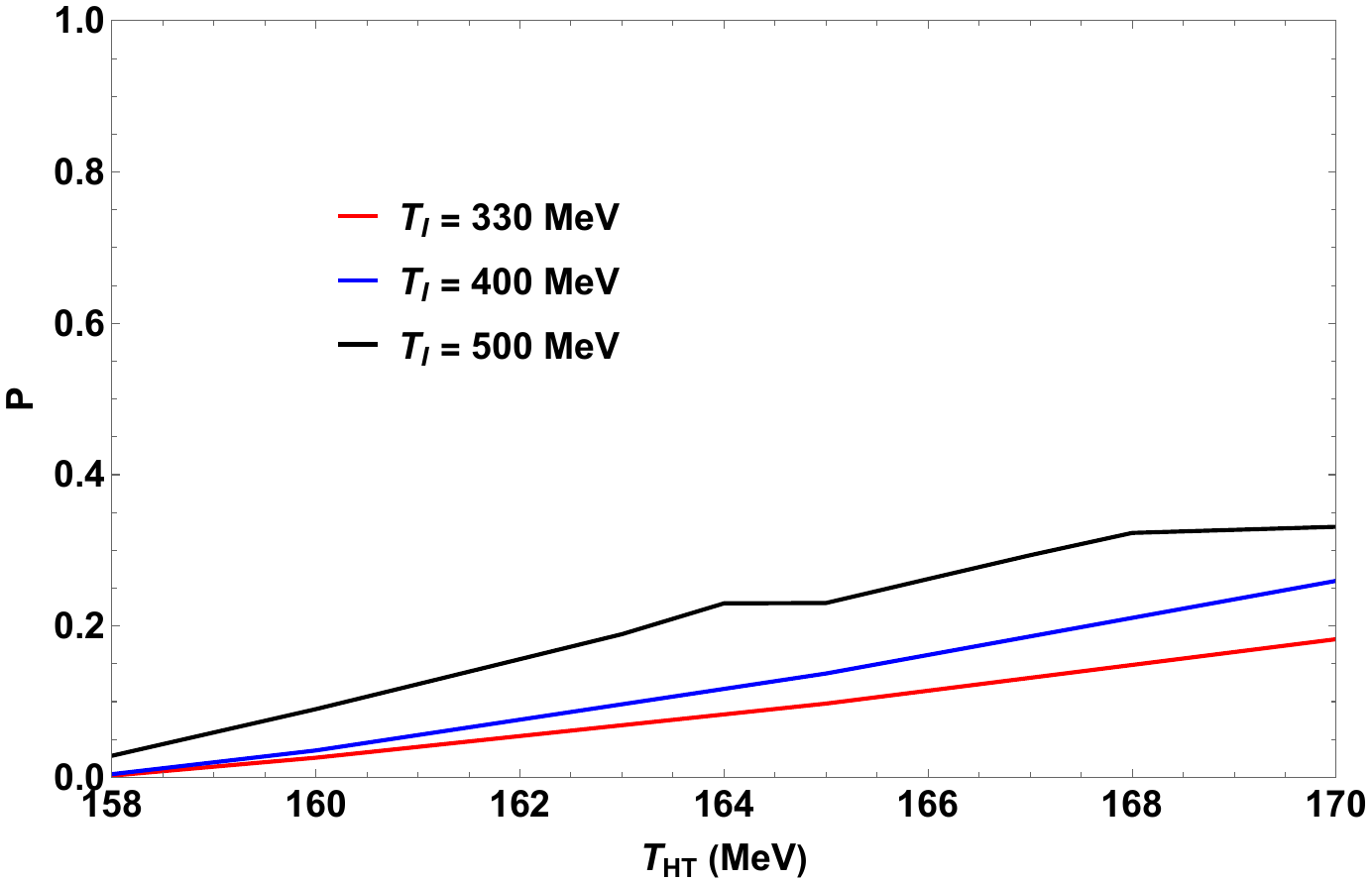}
  \caption{The probability that constituents of the hypertriton, the $\Lambda$, the proton and the neutron are inside the freeze-out boundary specified by the condition $T_F=156.5\, \text{MeV}$ as a function of the temperature of the center of mass with $\eta/s=0.12$. The likelihood that the constituents lie within the fireball increases with increasing initial central temperature.}
\label{Fig:PVsTCMWithtLC}
\end{figure}

A more useful measure of the tension between the assumptions of the SHM and the dynamics of the of the fireball is to look at the time-integrated, as opposed to the instantaneous probability of whether a hypertriton produced thermally inside the fireball can fit all of its components inside the fireball when the center of mass of the hypertriton is at a specific temperature. The results are shown in Fig.~(\ref{Fig:PVsTCMWithtLC}) for three different initial central temperatures. 

    From Fig.~(\ref{Fig:PVsTCMWithtLC}), we can see that the probability of the hypertriton having its center of mass at surface with $T=158\, \text{MeV}$ (that just within the freeze-out hypersurface) is almost negligible for $T_{I}=330\, \text{MeV}$ and $T_{I}=400\, \text{MeV}$, and less than $3\%$ for $T_{I}=500\, \text{MeV}$. The likelihood increases with the temperature of the center of the hypertriton,  $T_{\HT}$. Even when the initial central temperature of the fireball is as high as $500\, \text{MeV}$ and the hypertriton is located at a position where the local temperature is unrealistically large as 170 MeV, the probability is less than $33\%$. The probability($P$) that all of the constituents of the hypertriton are inside the fireball is lesser for a lower initial central temperature.

Viscous corrections could also play a role in determining the radial distances of the isotherms in the $\CM$ frame. We do not expect the bulk viscosity to play a substantial role. In general, the probability that all of the constituents of the hypertriton lie inside the fireball decreases with increasing $\eta/s$. The conjectured lower-bound on the $\eta/s$ value\cite{Kovtun:2004de}, therefore, also indicates a plausible upper-bound on  reasonable values for the likelihood of the hypertritons being inside the fireball given an initial central temperature for the Glauber profile. In our study, we found the variation of the likelihood with $\eta/s$ over the range $1-1.5$ in units of $1/(4\pi)$ to be negligible. Therefore, we present the results with $\eta/s=1.5/(4\pi)$, which we take as a reasonable estimate of $\eta/s$.

The obvious conclusion is that for reasonable choices of parameters, the hypertriton when located near enough to the freeze out hypersurface to be described by a hadron resonance gas with a density close to that seen in the SHM,  is typically too large to fit into the fireball.  This is a serious phenomenological problem for the SHM.  As will be seen in the following subsection, the large size of the hypertriton remains problematic given the temperature gradients in the fireball even when the hypertriton is contained within the fireball.  In fact, as discussed in the next subsection, the problem is worse than this indicates.  The reason is that even if all of the components of the hypertriton are contained in the fireball, the components may be located at temperatures above 170 MeV, where the HRG has become problematic.

\subsection{Effective temperature felt by the hypertriton's components }

In this subsection, we use the same sort of analysis as above, but, rather than focusing on whether or not all three constituents (proton, neutron and $\Lambda$) are within the hypertriton, we focus on those that have all three components contained within the fireball.  We do this because, even when all of the components are within the fireball, the hypertriton is large enough so that its components may well be at locations where the temperatures are high enough to be in serious phenomenological conflict with the SHM estimate that the hypertriton freezes out at 156.5 MeV. We will take a constituent to be in serious conflict when it is located at a place where the temperature is greater than 170 MeV. Recall that the yield of hypertritons predicted from a statistical hadronization model at $170\,\text{MeV}$ is about five times larger than the corresponding prediction at $156.5\, \text{MeV}$.

\begin{figure}[H]
  \centering
  \includegraphics[width=\textwidth]{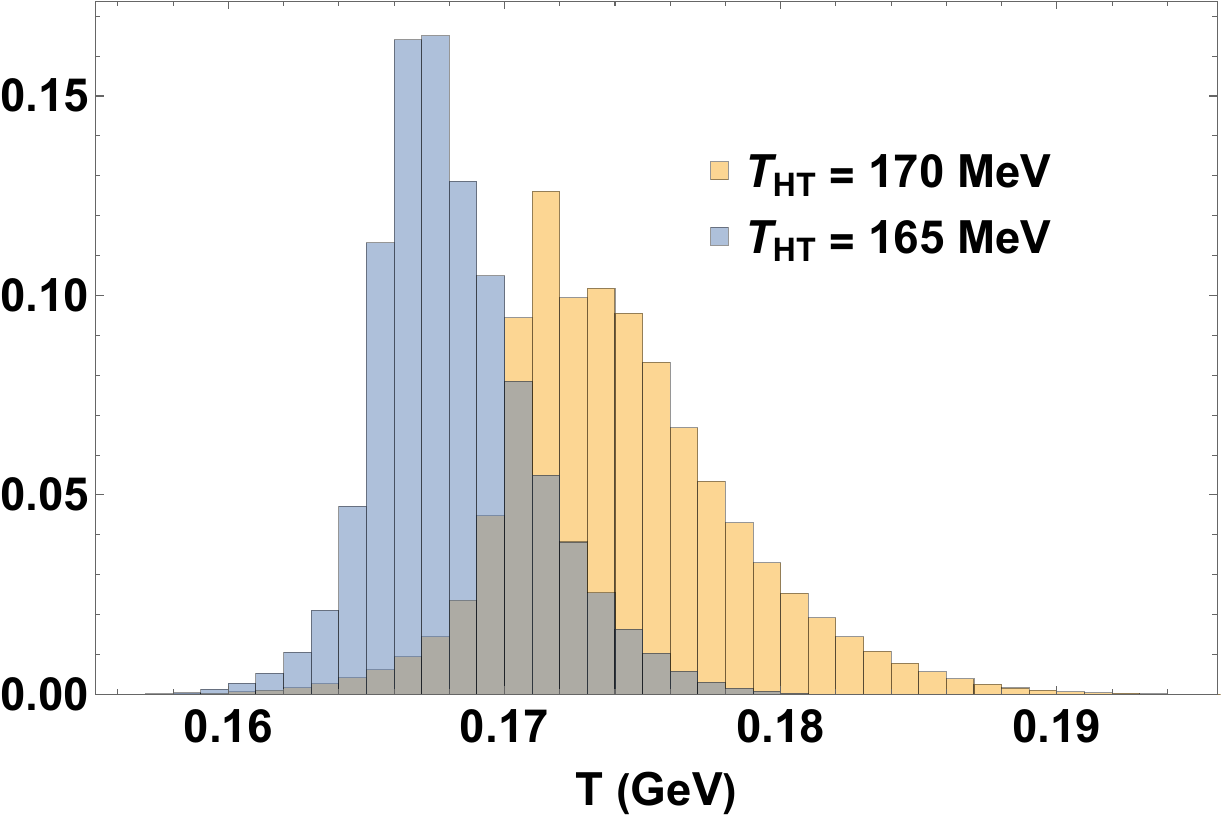}
  \caption{The plots show the normalized histogram for the largest of the three temperatures in the fluid at the location of the $p$, $n$ and $\Lambda$  among the events when the constituents of the hypertriton $\Lambda, n$ and $p$ are inside the fireball. Yellow and blue histograms corresponds to scenarios with temperature at the center of hypertriton equal to $170\, \text{MeV}$ and $165\, \text{MeV}$ respectively. The bin size for the histogram is $1\, \text{GeV}$ in temperature. The likelihood that all the hadronic constituents, considered point-like lie within the fireball with a given $T_{\HT}=170\, \text{MeV}$ is given in Fig.~(\ref{Fig:PVsTCMWithtLC}). $T_{I}\, =\, 500\, \text{MeV}$ and $\eta/s\,=\, 0.12$ for this hydro simulation. 
  }
\label{Fig:THT170TemDistribution}
\end{figure}

There is a probability distribution for the location of the constituents and associated with each location is a temperature.   Given that a hypertriton fits inside the fireball, we have computed the probability distribution (integrated over the full history of the fireball) for the largest of the
three temperatures in the fluid at the location of the $p$, $n$ and $\Lambda$. For this analysis, we assume that the initial central temperature is 500 MeV (our conservative estimate for an initial fireball central temperature).  These probability distributions depend on the position of the hypertriton and, as before, we will use the temperature at the center-of-mass of the hypertriton to label the position.  
These distributions are illustrated as histograms in Fig.~(\ref{Fig:THT170TemDistribution}) for two temperatures at the center of the hypertriton, 165 MeV and 170 MeV, where the histograms represent the probability distribution for the  temperature seen by the constituent located at the highest temperature.  
Fig.~(\ref{Fig:THT170TemDistribution}) shows the not so surprising fact that when the center of the hypertriton is located as far inside the fireball that the center is at temperature of 170 MeV, even when all components of the hypertriton are inside the fireball, statistically the fraction of situation in which none of the constituents is at a temperature that is greater than 170 MeV is quite small---about $10\%$.

\begin{figure}[H]
  \centering
\includegraphics[width=\textwidth]{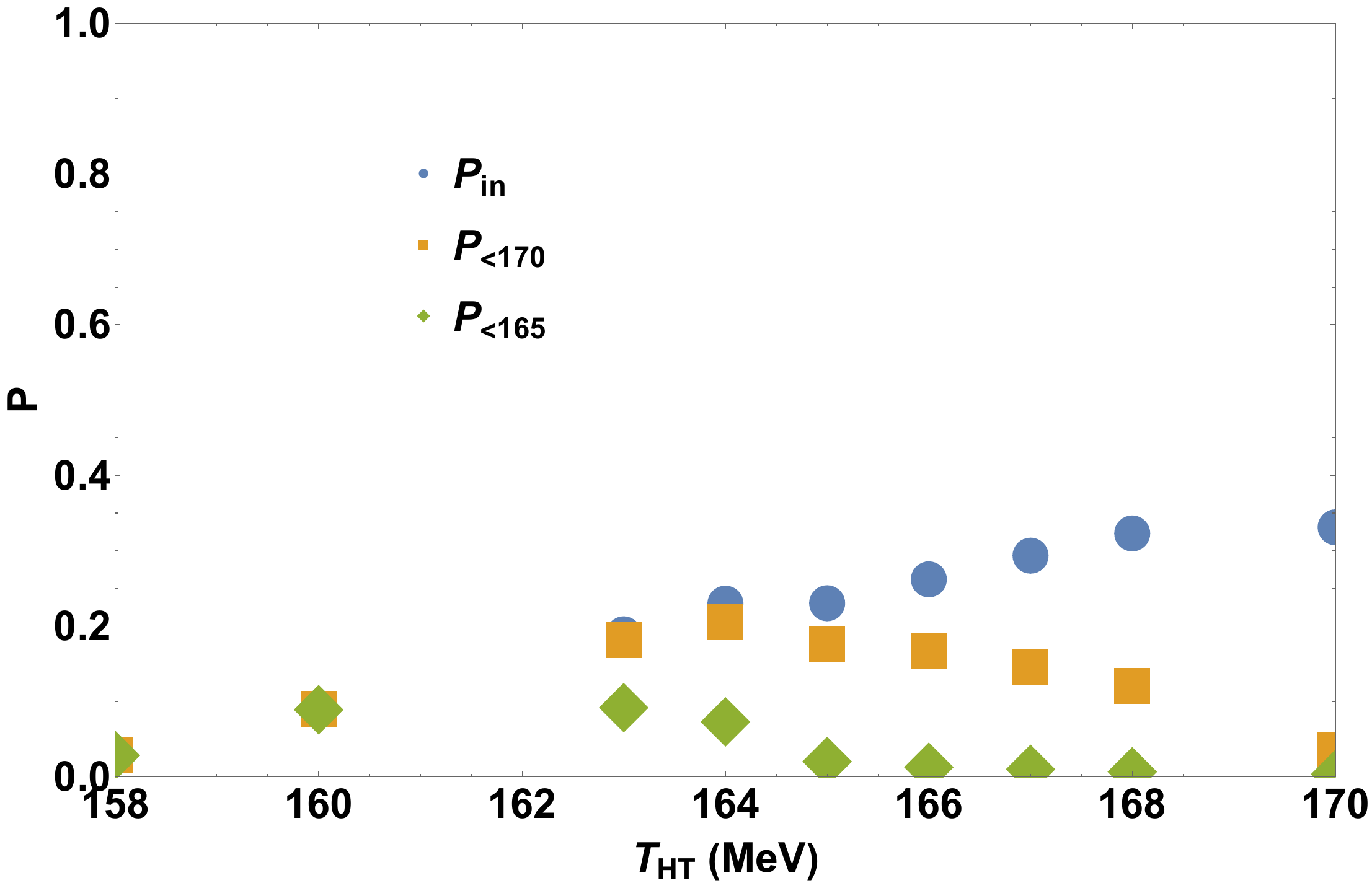}
  \caption{
  The yellow squares indicate the probability that all the constituents lie within the shell marked by the fireball boundary and the isotherm of $T=170\, \text{MeV}$.    The green squares indicate the probability that all the constituents lie within the shell marked by the fireball boundary and  isotherm of $T=165\, \text{MeV}$.  For comparison, the blue circles show the probability that all constituents are within the fireball.
  }
\label{Fig:CDF}
\end{figure}

Of course, for the SHM to be justified in a natural way for hypertritons, one needs to have hypertritons formed and equilibrated just inside the freeze out surface.  This is sensible only if the center of the hypertriton and all of its hadronic constituents are both inside the fireball and at a temperature close to the chemical freeze out temperature.   For the purposes of an estimate for the probability that this occurs, we have taken 170 MeV, an extremely conservative value, as sufficiently close to the chemical freeze out temperature.  Fig.~(\ref{Fig:CDF}) shows just how rare this.  For all cases, it is less than 25\% and is typically much smaller. This would be enough to spoil the phenomenological success of the SHM. 
Moreover, if one takes a more reasonable upper bound for ``close'' to the freeze out temperature, than 170 MeV, the probabilities would have dropped markedly. Fig.(\ref{Fig:CDF}) indicates that if one uses 165  MeV as  the upper bound for sufficiently close, the maximum probability drops to about 10\% and is typically much smaller.

The conclusion of this analysis is simple: given the spatial gradients of the isotherms during the time of evolution, the hadronic components of the hypertriton are spread out to such a degree that they and the center of the hypertriton are unlikely to be inside the fireball and at a temperature sufficiently close to the chemical freeze-out to be phenomenologically relevant.  Thus, it is puzzling as to why the SHM gets the yield of hypertritons approximately correct.

\section{Can a compact colorless quark droplets at freeze-out evolve into the hypertriton state? }

\label{last}

Recall that  
Ref.~\cite{Andronic:2017pug} while advocating for the Statistical Hadronization Model  as a way to decode the phase structure of QCD, also observed  that it was remarkable that the hypertriton yields agreed so well with the model given the large size of the hypertriton relative to the fireball.  The paper speculated that the agreement might require a production mechanism in which involving a primary ``compact" hypertriton state---a colorless quark droplet---formed  at or near freeze out temperatures, which  subsequently evolves into the weakly bound hypertriton state seen asymptotically.  Ref.~\cite{Andronic:2017pug} suggests such droplets should have a ``lifetime'' of at least 5 fm (and by the energy-time uncertainty relation energies of less than 40 MeV). 

We note here that this speculation, even if correct,  does not resolve the hypertriton puzzle by itself without further dynamical assumptions.  Of course, it is logically possible that compact quark droplets with quantum numbers of the hypertriton do form near the freeze out surface.  A key difficulty is that quantum mechanically such quark droplets must correspond to superpositions of physical states with the appropriate quantum numbers.  Thus, the ``lifetime'' of the droplet corresponds to the time over which the the various physical components making up the droplet dephase to the point that they can effectively be  distinguished.  For the case of the hypertriton quantum number, the relevant physical states are the hypertriton, deuteron plus $\Lambda$ scattering states and three-body proton, neutron plus $\Lambda$ scattering states.  Such droplets evolve into hypertriton only a certain small fraction of the time and into various scattering states, the rest of the time.  Suppose that the  fraction of such droplets that ultimate evolve into a hypertriton and not a scattering state is denoted $f$.  Since the object needs to be very compact compared to the hypertriton to evade the difficulties noted in this paper, one would require that $f \ll 1$.  On the other hand, for the SHM to  correctly predict the yield of hypertritons based on a droplet picture, the system would need to dynamically produce a density of droplets near the freeze out surface  that is $1/f$ times the would-be density of equilibrated hypertritons.  {\it A priori}, there is no obvious reason why such a density of droplets should  be produced and without a natural explanation for such a density, the hypertriton puzzle remains unresolved.

\begin{acknowledgments}
This work has been supported by the DOE grant DE‐FG02‐93ER40762. The authors acknowledge useful discussions with Eric Braaten.
\end{acknowledgments}

\appendix

\section{Hydrodynamic Equation of State used}
\label{appendeos}

\begin{figure}[H]
  \centering
\begin{minipage}{0.45\textwidth}
  \includegraphics[width=\textwidth]{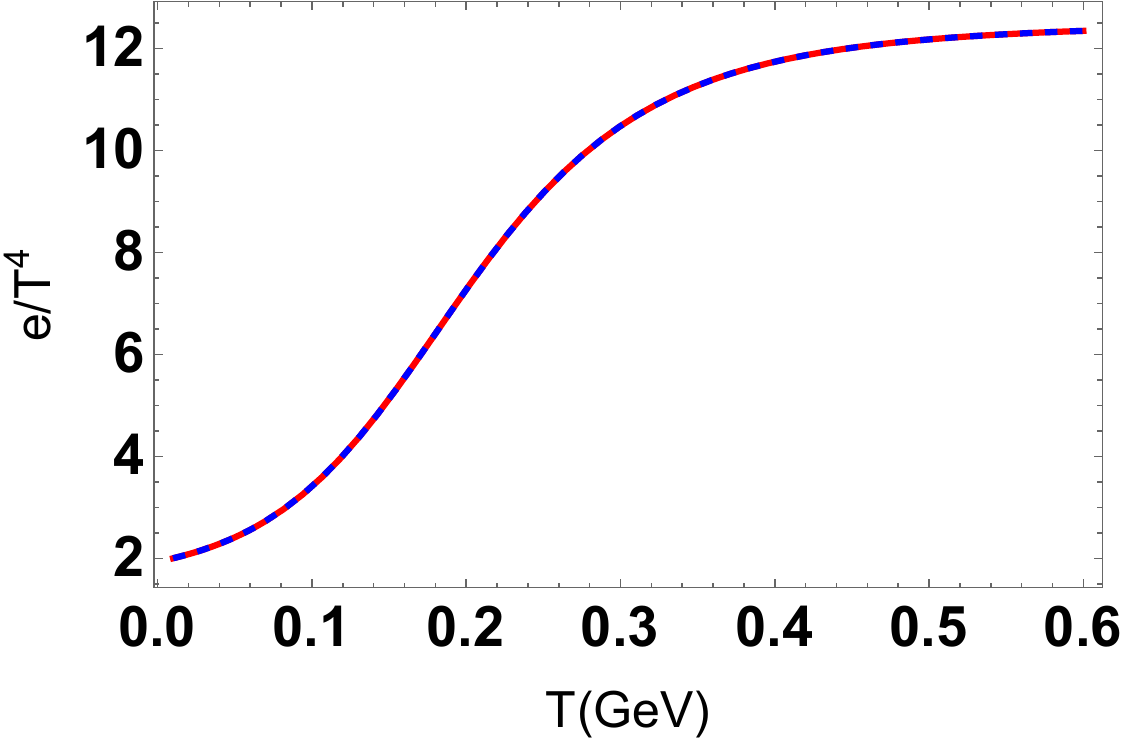}
  \end{minipage}
  \begin{minipage}{0.45\textwidth}
  \includegraphics[width=\textwidth]{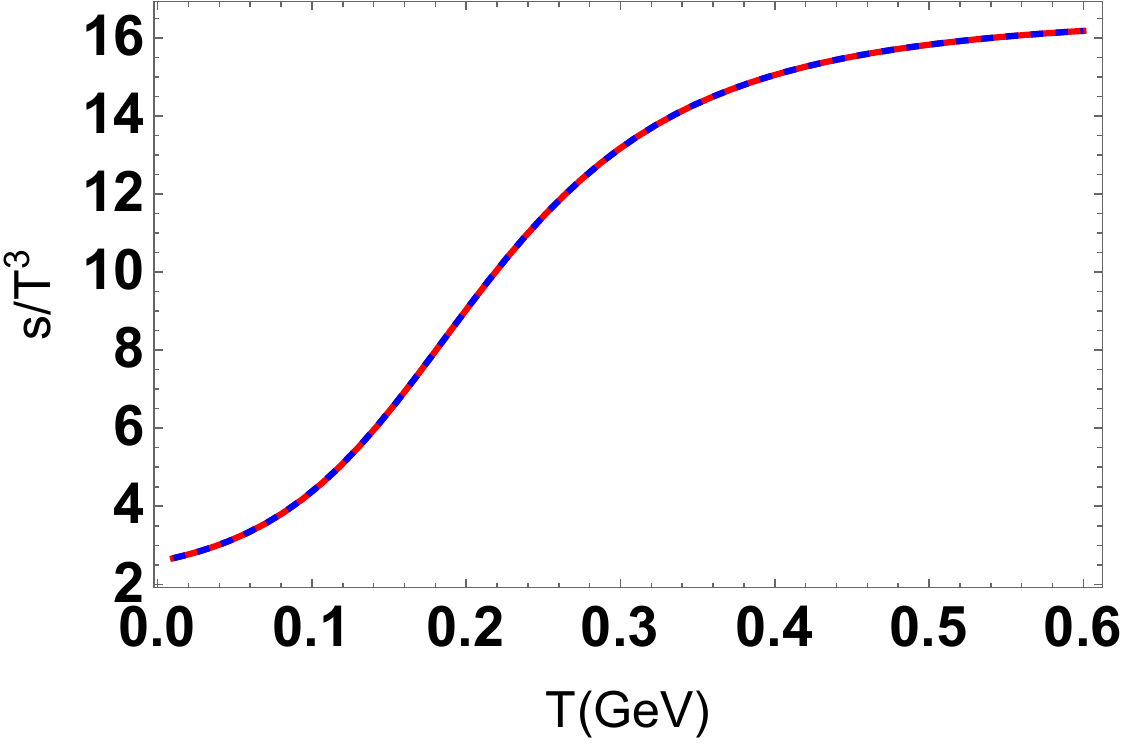}
  \end{minipage}
  \caption{The energy density and the entropy density, normalized by appropriate powers of temperature as a function of the temperature, $T$.}
\label{Fig:eAnds}
\end{figure}
\begin{figure}[H]
  \centering
\begin{minipage}{0.45\textwidth}
  \includegraphics[width=\textwidth]{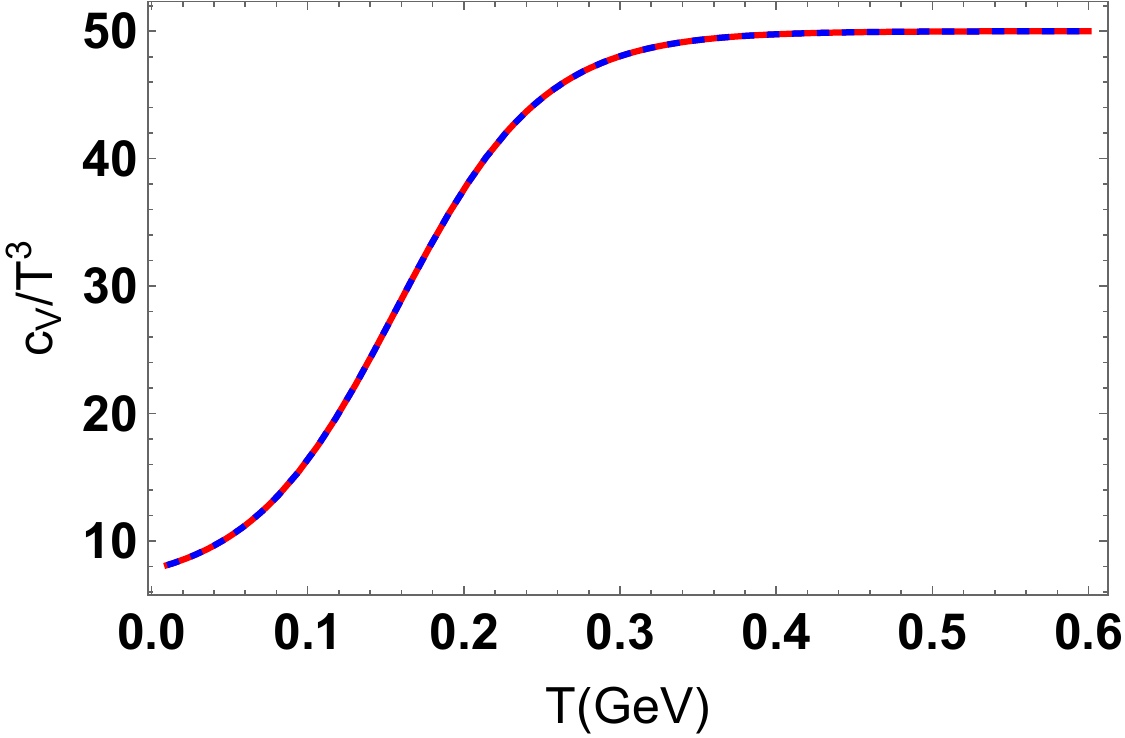}
  \end{minipage}
  \begin{minipage}{0.45\textwidth}
  \includegraphics[width=\textwidth]{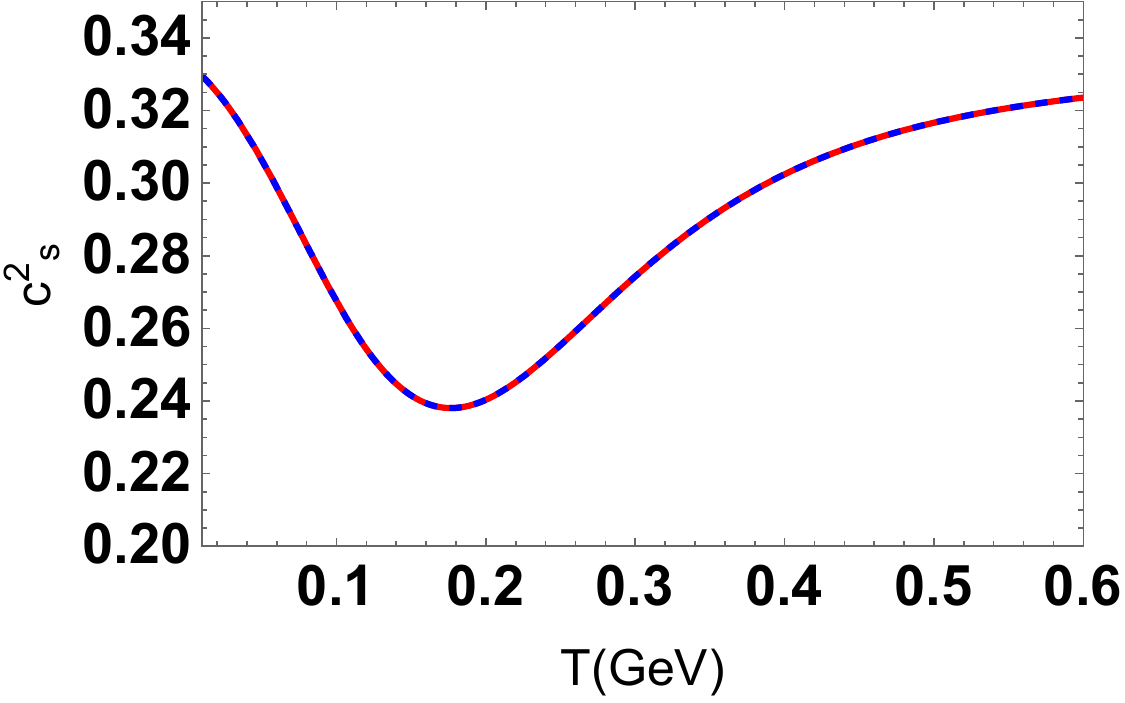}
  \end{minipage}
  \caption{The specific heat at constant volume and the speed of sound squared, normalized by appropriate powers of temperature as a function of the temperature, $T$.}
\label{Fig:cvAndcssquared}
\end{figure}

The equation of state used in this work was used before in Ref.~\cite{Rajagopal:2019xwg,Pradeep:2022mkf}. Unlike there, the effects of a critical point are switched off here as we are interested in a phenomenological EoS suited for high energies.

\bibliographystyle{apsrev4-1}
\bibliography{refs}

\end{document}